\documentstyle[aps,pra,epsfig,manuscript]{revtex}

\begin{document}
\title{Generation of higher-order atomic dipole squeezing in a \\  high-Q
micromaser cavity:  VIII. multi-photon interaction} 
\author{Rui-Hua Xie$^{a}$\footnote{Email: rhxie@titan.nist.gov; Phone: 
(301) 975-5159; Fax: (301) 990-1350} and Qin Rao$^{b}$}
\address{$^{a}$Department of Chemistry, Queen's University, Kingston,
ON K7L 3N6, Canada;\\
$^{b}$Department of Engineering Physics, Queen's University, Kingston,
ON K7L 3N6, Canada}
\date{\today}
\maketitle
\begin{abstract}
In our preceding serial works,  we have investigated the generation of 
higher-order atomic dipole squeezing (HOADS) in a high-Q micromaser 
cavity, discussing the effects of dynamic Stark shift, atomic damping, 
atomic coherence and nonlinear one-photon processes and different initial 
states (for example, correlated and uncorrelated states, superposition 
states, squeezed vacuum). In this paper, we continue to study HOADS in 
a high-Q micromaser cavity, but consider that the atom interacts with 
the optical field via  a multi-photon transition process and that 
the initial atom is  arbitrarily prepared. For a vacuum initial field, 
we demonstrate that HOADS cannot occur if the atom is initially 
prepared  in a chaotic state and that a coherent atomic state 
generates less efficient and stable HOADS than an arbitrary one. 
It is found that large detuning may lead to enhanced and strong HOADS.

\noindent
{\bf PACS (numbers)}: 42.50.Dv, 42.50.Lc, 32.80.-t


\end{abstract}

\newpage

\begin{center}{\bf 1. INTRODUCTION}  \end{center}
 
In 1926, Heisenberg\cite{heisen} proposed a fundamental, general 
principle of nature, known as Heisenberg uncertainty principle. 
Although this  principle  is of no consequence for the macroscopic world, 
it does play an important role in dealing with problems met in the 
microscopic world (for example, atom, molecule, nanocrystals, solid,
 and so on). It was Kennard\cite{kennard} who presented us the first 
example of nonclassical states, called {\sl squeezed state}, in which 
the quantum fluctuations in a dynamical observable may be reduced below 
the standard quantum limit at the expense of  increased fluctuations 
in its canonical conjugated one without violating the Heisenberg uncertain 
principle. Furthermore,  Pleba\'{n}ski\cite{pleb} made an important 
contribution to the theory of squeezed states in 1956. 
The exciting and constructive experimental results of the 
first successful generation and detection of squeezed states was 
reported in the middle  of the 1980's\cite{res85}. Since then, the 
squeezed states have been  extensively studied, both theoretically 
and experimentally
\cite{kw87,special92,xie95,xie961,xie962,xie963,xie964,xie965,gb97,af98,book00,xie2001,xie2002}, 
due to their potential applications in quantum communication, high-resolution 
laser spectroscopy measurement, gravity wave detection and quantum 
information theory  (quantum teleporation, dense coding,  cryptography, 
quantum nondemolition measurement and  power-recycled interferometer). These 
extensive studies have shown that a lot of nonlinear optical systems could 
generate squeezed states for the field and atom, for example, in two-photon 
laser, parametric  amplifiers,four-wave mixing, resonance fluorescence, 
Rydberg atom maser and cooperative Dicke system. A general relationship 
between field and atomic dipole squeezing has even been established under 
different initial conditions for both the field and atom
 \cite{kw87,xie961,xie963,xie964,xie965}.  

Recently, the rapid development of techniques for making higher-order 
correlation measurements has resulted in an increased interest in 
generating another kind of nonclassical states, known as  higher-order 
squeezed states \cite{r17}. In these states,  higher-order quantum 
fluctuations in one quadrature of the field or atomic dipole could 
be reduced without violating higher-order uncertainty relations\cite{r17}. 
It has been predicted that  higher-order squeezed states for the radiation 
In 1997,  one of us, Xie,  with his colleagues\cite{ref24} introduced the 
concept of higher-order atomic dipole squeezing (HOADS)
and applied it to high-Q micromaser cavities. In our serial 
works\cite{ref244,ref245,xie021,xie022,xie023,xie024}, we have discussed 
thoroughly the important connection of HOADS with the second-order field 
and atomic dipole squeezing (ADS), the effects of a  nonlinear one-photon 
process and dynamical Stark shift in two-photon processes on HOADS, and 
different initial conditions regarding both the atomic and field (for 
example, uncorrelated and correlated coherent states, superposition states, 
squeezed vacuum). These extensive studies have made an important contribution 
to the theory of higher-order squeezed states and  provided us an approach 
to extracting information efficiently from an optical signal by higher-order 
correlation measurements. However, the initial conditions are more or less 
individualized, and  only one- and two-photon transitions have been 
investigated. Hence, it is our purpose in this paper to treat this problem 
in a more general way. As a first step, we consider that the initial atom 
is arbitrarily prepared and that the atom interacts with the field via a 
multi-photon transition process. In this paper, we use the two-level 
multi-photon Jaynes-Cummings model \cite{xie962,xie968,m27} and utilize 
the density matrix method which can describe very well arbitrary atomic-field 
states such as a mixed state. Their actual squeezing conditions and behaviours 
were carefully discussed. 
 
This paper is organized as follows. In section 2, we describe
the theoretical model of multi-photon interaction between a cavity 
field and a two-level atom and introduce the concept of HOADS.  In 
section 3, assuming that the atom is  initially prepared in an 
arbitrarily  state, we  investigate the generation of HOADS in the 
multi-photon Jaynes-Cummings model. A summary is given in the last 
section.

\begin{center}{\bf 2. THEORY}\end{center}

Here we consider the interaction of a two-level atom with a single-mode 
quantized field involving the emission or absorption of multi-photons 
per atomic transition. The Hamiltonian of this system can be written 
as follows
\begin{equation}
H = \Omega a^{\dagger}a + \omega S_{z} + \epsilon (a^{\dagger \xi} 
+ a^{\xi})( 
S_{-}+S_{+})
\end{equation}
which is the multi-photon Jaynes-Cummings model without the rotating-wave 
approximation (RWA)\cite{xie962,xie968}. If the RWA is explicitly used, 
then we have 
\begin{equation}
H = \Omega a^{\dagger}a + \omega S_{z} + \epsilon(a^{\dagger\xi}S_{-} 
+ a^{\xi}S_{+})
\end{equation} 
which is the multi-photon Jaynes-Cummings model within the RWA\cite{m27}. 
$S_{z}$ and $S_{\pm}$ are operators of
the atomic pseudospin inversion and transition, respectively,
and satisfy the commutation \cite{xie95}: 
$[S_{+}, S_{-} ] = 2 S_{z}$ and 
$[S_{z}, S_{\pm} ] = \pm S_{\pm}$. 
$\omega$ is the transition frequency for the atom.
$a^{\dagger}$ and $a$ are the creation and
annihilation operators for the photons with the frequency $\Omega$,
 which obey the boson operators'
commutation relations, $[a, a^{\dagger}]=1$.
$\epsilon$ is the coupling constant between the atom and the radiation field, 
and $\xi$ is the absorbing or emitting photon numbers per atomic transition. 
Throughout we employ the unit with $\hbar=c=1$. 
 
In order to investigate the squeezing properties of the
atomic dipole variables, we follow the standard
procedure of defining the slowly varying operators\cite{xie95}
\begin{eqnarray}
S_{x}&=&\frac{1}{2}\left[ S_{+} e^{-i\omega t}+S_{-} e^{i\omega t}\right],\\
S_{y}&=&\frac{1}{2i}\left[S_{+} e^{-i\omega t}-S_{-} e^{i\omega t}\right],
\end{eqnarray}
where $S_{x}$ and $S_{y}$, in fact, correspond to the dispersive and
absorptive components of the slowly varying atomic dipole
\cite{rref28}, respectively. One can easily show that the above
operators obey the commutation relation, $[S_{x}, S_{y}] = i S_{z}$.
Correspondingly, we found the higher-order uncertainty relation\cite{ref24}
concretely given by
\begin{equation}
(\Delta S_{x})^{P}(\Delta S_{y})^{P} \ge \frac{1}{4}
\left|[(\Delta S_{x})^{P/2}, (\Delta S_{y})^{P/2}]\right|^{2}.
\end{equation}
For a two-level atom, we have 
\begin{equation}
(\Delta
S_{j})^{P}=2^{-P}+\sum_{k=2,4,...}^{P}(C_{P}^{k}-C_{P}^{k-1})2^{k-P}<S_{j}>^{k}
\ \ \ \ \ \ (j=x, y),
\end{equation}
\begin{equation}
[(\Delta S_{x})^{P/2}, (\Delta
S_{y})^{P/2}]=i<S_{z}>\sum_{m,n}^{P/2-1}C_{P/2}^{m}
C_{P/2}^{n}2^{2-P+m+n}<S_{x}>^{m}<S_{y}>^{n},
\end{equation}
where $P/2\ge 1$ is an integer.
If $P/2$ is even (odd), then $m$ and $n$ are odd (even) numbers.
It is convenient that we define the following functions
\begin{eqnarray}
F_{1}(P)&=&(\Delta S_{x})^{P}-\frac{1}{2}\left| [(\Delta S_{x})^{P/2},
(\Delta S_{y})^{P/2}]\right|, \\ 
F_{2}(P)&=&(\Delta S_{y})^{P}-\frac{1}{2}\left| [(\Delta S_{x})^{P/2},
(\Delta S_{y})^{P/2}]\right|.
\end{eqnarray}
 Then, higher-order quantum fluctuations in the component $S_{x}$ (or $S_{y}$) 
of the dipole  are squeezed if $F_{1}<0$ (or $F_{2}<0$). This is the general 
definition of HOADS \cite{ref24}.

\begin{center}{\bf 3. Results}\end{center}

We denote $\mid n\rangle$ as the Fock state of the radiation field 
and $\mid +\rangle$ and $\mid -\rangle$ as the excited and 
ground states of the two-level atom, respectively.  Using the 
standard bare-state procedure introduced by Xie {\sl et al.}
\cite{x1997}, we can calculate the density operator of the atom-field 
coupling system at time 
$t$ with an arbitrary initial condition $\rho(t=0)$ by
\begin{equation}
\rho(t)=U(t)^{\dagger}\rho(t=0)U(t)\equiv\sum_{ij}\rho_{ji}(0) 
e^{-i(E_{i}-E_{j})t}\mid j\rangle\langle i\mid
\end{equation}
where $\rho_{ji}(0)=\langle j\mid\rho (0)\mid i\rangle$. 
$\mid i\rangle$ and $E_{i}$ are the $i$th eigenstate and 
its corresponding eigenvalue, respectively,  and $U(t)=\exp(-iHt)$ 
is the unitary evolution operator.  Therefore, the expectation value of any physical operator $O$  
at time $t$  can be arrived at through
\begin{eqnarray}
\langle O(t)\rangle &=& Tr[\rho(t)O(0)] = 
\sum_{i}\langle i\mid\rho(t)O(0)\mid i\rangle\nonumber\\
&=&\sum_{ij}\rho_{ij}(t)\langle j\mid O(0)\mid i\rangle=
\sum_{n=0}^{\infty}\sum_{m=0}^{\infty}\sum_{\alpha=1}^{2}\sum_{\beta}^{2}
\rho_{n\alpha,m\beta}(t)\langle\Psi_{m,\beta}\mid O(0)\mid
\Psi_{n,\alpha}\rangle, 
\end{eqnarray}
where $\mid i\rangle=\sum_{n=0}^{\infty}\sum_{\alpha=1}^{2}\mid \Psi_{n,\alpha}\rangle$. Here 
$\mid\Psi_{n,\alpha}\rangle$ denotes the dressed state 
with $n$ being the photon number of the optical field. The wave function 
$\mid\Psi_{n,\alpha=1,2}\rangle$  can be expressed by 
\begin{eqnarray}
\mid\Psi_{n,1}\rangle&=&(\mid+,n\rangle+\mid -,n+p\rangle)/\sqrt{2},\\
\mid\Psi_{n,2}\rangle&=&(\mid+,n\rangle-\mid -,n+p\rangle)/\sqrt{2} 
\end{eqnarray}
with the corresponding eigenenergies 
\begin{eqnarray}
E_{n,1}&=&(n+\xi/2)\omega+\epsilon\Pi_{n},\\
E_{n,2}&=&(n+\xi/2)\omega-\epsilon\Pi_{n}, 
\end{eqnarray}
where $\Pi_{n}$ is defined by
\begin{equation}
\Pi_{n}=\sqrt{\frac{(n+\xi)!}{n!}}.
\end{equation}
As usual, we assume that the initial density matrix $\rho(t=0)$ can 
be decomposed into its atomic and field parts, i.e., 
$\rho(0)=\rho_{a}(0)\otimes\rho_{f}(0)$.  In this paper, we 
assume that the initial atomic state is arbitrarily 
prepared, i.e., 
\begin{equation}
\rho_{a}(0) =\sin^{2}\theta\mid +\rangle\langle +\mid + 
 \cos^{2}\theta\mid -\rangle\langle -\mid+ 
\eta\left(e^{-i\phi}\mid -\rangle\langle +\mid
+e^{i\phi}\mid +\rangle\langle -\mid\right)
\end{equation} 
with 
\begin{equation}
\eta^{2} \le \sin^{2}\theta\cos^{2}\theta\le \frac{1}{4},
\end{equation}
and the field is in its vacuum state, i.e., 
\begin{equation}
\rho_{f}(0)=\mid 0\rangle\langle 0\mid.
\end{equation} 
Then, we have arrived at the  expectation values of 
$S_{x}$, $S_{y}$  and $S_{z}$
\begin{eqnarray}
<S_{x}(t)>
&=&\eta\left\{\cos(\Gamma\tau)\cos\left(\frac{\Delta}{2}\tau+\phi\right)-
\Delta\frac{\sin(\Gamma\tau)}{2\Gamma}\sin\left(\frac{\Delta}{2}\tau+\phi\right)
 \right\},\\
<S_{y}(t)>
&=&\eta\left\{\cos(\Gamma\tau)\sin\left(\frac{\Delta}{2}\tau+\phi\right)-
\Delta\frac{\sin(\Gamma\tau)}{2\Gamma}\cos\left(\frac{\Delta}{2}\tau+\phi\right)
 \right\},\\
<S_{z}(t)>
 &=&\frac{1}{2}\left\{\sin^{2}\theta\frac{\Delta^{2}/4+\xi!
\cos(2\Gamma\tau)}{\Gamma^{2}}-\cos^{2}\theta\right\},
\end{eqnarray}
where $\tau=\epsilon t$ is the scaled interaction time, 
 $\Delta=(\xi\Omega-\omega)/\epsilon$ is the scaled detuning, 
$\Gamma=\sqrt{\Delta^{2}/4+\xi !}$ is the scaled 
Rabi frequency and $\phi$ is the relative phase 
between the excited state $\mid +\rangle$ and the ground state  
$\mid -\rangle$ of the two-level atom. If we take $\eta=\sin\theta\cos\theta$, 
the above results are in agreement with those of Xie {\sl et al.}
\cite{xie962,xie968}. 

If the two-level atom is initially prepared in a chaotic state 
(i.e., $\eta=0$), one can easily see that the second- and higher-order 
quantum fluctuations in $S_{x}$ or $S_{y}$ cannot be squeezed.  
In the case of pure initial atomic states, which still keep all the
relevant physical features, we have $\eta=\mid\sin(2\theta)\mid/2$ and 
find that no HOADS occurs for a completely excited or ground atom, but 
HOADS could appear for a coherent atom as shown below.

One interesting result is the so-called population trapping 
\cite{cirac}, in which the expectation value of $S_{z}$ takes on a 
steady value, i.e., $d <S_{z}(t)>/dt=0$. This results in 
$\sqrt{\xi !}\tau=k\pi$ or $(k+1/2)\pi$ (k=0, 1, 2, ...). 
Obviously, for $\sqrt{\xi !}\tau=(k+1/2)\pi$, the second- and higher-order
quantum fluctuations in $S_{x}$ or $S_{y}$ cannot be squeezed.
However, for $\sqrt{\xi !}\tau=k\pi$, HOADS occurs for proper choice 
of the parameters $\eta$, $\theta$ and $\phi$. Details are given below.

In Fig.1, we show a 3D contour plot for the function $F_{1}(P=6)$, 
$\theta$ and $\eta$ for $\phi=(k+1/2)\pi$ (k=0,1,2,...) in the case 
of resonance.  We find that for $0<\eta\le 0.5$ the sixth-order 
quantum fluctuations in $S_{x}$ can be squeezed by properly choosing 
the parameter $\theta$. Especially, when $\eta=0.5$, HOADS appears 
almost in the whole range of $\theta$ except a small range 
around $\theta=\pi/4$ or $\theta=3\pi/4$. Also it is obvious that 
the pure atomic state with  $\eta=\mid\sin(2\theta)\mid/2$ generates 
less HOADS than the atomic state with $0<\eta\le 0.5$. However, as 
we vary the phase to $\phi=k\pi$ (k=0,1,2,...), as shown in Fig.2, there 
exists only a small region of HOADS in the $\eta$-$\theta$ parameter space. 
This implies that HOADS is much sensitive to the relative phase 
between the excited and ground states of the two-level atom. 
 
In Fig.3 and 4, we show 3D contour plots for the function $F_{1}(P=6)$,
parameter $\theta$ and phase $\phi$ for $\eta=\mid\sin(2\theta)\mid/2$ 
and $\eta=0.5$, 
respectively, for a resonance transition (i.e., $\Delta=0$).
 HOADS patterns are clearly shown in the $\theta$-$\phi$ parameter space. 
We find that HOADS is much sensitive to the phase for an initial pure atomic 
state, whereas it is relatively stable with respect to the phase for 
an arbitrary state with $\eta=0.5$.  This implies that an  arbitrary atomic 
state could generate much efficient and stable HOADS than a pure one. 

In Fig.5, we show the time evolution of the function $F_{1}(P=6)$ 
for $\theta=\pi/2$, $\eta=0.5$, $\phi=k\pi$ in the case of resonance. 
We notice that increasing the photon number $\xi$ results in a decrease of the 
HOADS duration, which is inversely proportional to 
$\sqrt{\xi!}$. The HOADS frequency is directly proportional to
$\sqrt{\xi!}$. 
 
In Fig.6, we show the time evolution of the function $F_{1}(P=6)$
for $\theta=\pi/2$, $\eta=0.5$ and $\phi=k\pi$ and $\Delta=0, 1, 5$ 
in the case of one-photon transition.  It is seen that the duration, 
period and strength of HOADS are much 
sensitive to the detuning. Especially, for a large detuning, 
for example $\Delta=5$, we observe enhanced and strong HOADS. Similar 
results, as shown in Fig.7,  are obtained for three-photon transition 
processes.  These results show that large detuning can generate 
efficient HOADS than the resonance case. 

Finally, it would be interesting to examine the relation between 
the second-order ADS (SOADS)  and HOADS. In Fig.8 and 9, we show 3D contour 
plots for the function $F_{2}(P=2)$, $\theta$ and $\phi$ for 
$\eta=\mid\sin(2\theta)\mid/2$ and $\eta=0.5$, respectively, for a 
resonance transition.  Comparing Fig.8 and 9 with Fig.3 and 4,
 respectively, we find  that there exist additional 
$\theta$-$\phi$ parameter regions where 
HOADS could occur but SOADS could not. There are also more 
HOADS dips than SOADS.  For an arbitrary  atomic  state with 
$\eta=0.5$, we see that  HOADS with respect to the fluctuation of 
the phase $\phi$ is much stable than SOADS. Certainly, as demonstrated 
in our previous serial works 
\cite{ref244,ref245,xie021,xie022,xie023,xie024}, there exist
common $\theta$-$\phi$ parameter regions where both HOADS and SOADS 
could be generated in the mean time. 

\begin{center}{\bf 4. SUMMARY}\end{center}

In summary, we study HOADS in a high-Q micromaser cavity by considering 
a two-leve atom interacting with an optical field through a multi-photon
transition. Assuming that the initial atom is arbitrarily prepared and 
the field is initially in a vacuum state, we demonstrate that  HOADS 
cannot appear if the atom is initially prepared in a chaotic state 
and that a coherent atomic state could generate less efficient and 
stable HOADS than an arbitrary one. It is found that large detuning 
may lead to enhanced and strong HOADS.

\begin{center}{\bf ACKNOWLEDGEMENT}\end{center}    

We would like to thank Jianing Colin for some important contribution 
and constructive conversation and encouragement. One of us (Q. R.) 
acknowledges the Reinhart Fellowship provided by Queen's University 
for financial support.

\newpage

\begin{center}{\Large\bf Caption of Figures}\end{center}

\

{\bf FIG.1}: The 3D contour plot of function $F_{1}(P=6)$,
$\eta$ and $\theta$ for $\Delta=0$,
$\sqrt{\xi!}\tau=k\pi$ and $\phi=(k+1/2)\pi$ (k=0,1,2,...),
where the dotted line is for $F_{1}(P=6)=0$,
the dotted-dashed lines for $F_{1}(P=6) =0.01$, 
and the dashed lines for $F_{1}(P=6) < 0$ in the range 
of -0.045 to -0.005 with an interval of 0.005. The solid 
line is for $\eta=\mid\sin\theta\cos\theta\mid$.

\

{\bf FIG.2}: The 3D contour plot of function $F_{1}(P=6)$, 
$\eta$ and $\theta$ for $\Delta=0$, 
$\sqrt{\xi!}\tau=k\pi$ and $\phi=k\pi$ (k=0,1,2,...), 
where the dotted line is for $F_{1}(P=6)=0$, 
the dotted-dashed lines for $F_{1}(P=6) > 0$ in the range of 
0.01 to 0.05 with an interval of 0.01, and the dashed lines for 
$F_{1}(P=6) < 0$ in the range of -0.06 to -0.005 with an interval 
of 0.005. The solid line is for $\eta=\mid\sin\theta\cos\theta\mid$.

\

{\bf FIG.3}: 
The 3D contour plot of function $F_{1}(P=6)$,
$\phi$ and $\theta$ for $\Delta=0$,
$\sqrt{\xi!}\tau=k\pi$ and $\eta=\mid\sin\theta\cos\theta\mid$,
where the dotted line is for $F_{1}(P=6)=0$,
the dotted-dashed lines for $F_{1}(P=6) > 0$ in the range of
0.01 to 0.05 with an interval of 0.01, and the solid lines for
$F_{1}(P=6) < 0$ in the range of -0.06 to -0.001 with an interval
of 0.001.

\
 
{\bf FIG.4}: 
 The 3D contour plot of function $F_{1}(P=6)$,
$\phi$ and $\theta$ for $\Delta=0$, 
$\sqrt{\xi!}\tau=k\pi$ and $\eta=0.5$, 
where the dotted line is for $F_{1}(P=6)=0$, 
the dotted-dashed lines for $F_{1}(P=6) > 0$ in the range of
0.01 to 0.05 with an interval of 0.01, and the solid lines for
$F_{1}(P=6) < 0$ in the range of -0.06 to -0.001 with an interval
of 0.001.

\

{\bf FIG.5}: Time evolution of the function $F_{1}(P=6)$ for 
$\eta=0.5$, $\theta=\pi/2$, $\Delta=0$ and $\phi=k\pi$ 
(k=0,1,2,...): (a) $\xi=1$ (solid line); (b) $\xi=4$ (dotted-dashed 
line).

\

{\bf FIG.6}: Time evolution of the function $F_{1}(P=6)$ for
$\eta=0.5$, $\theta=\pi/2$, $\phi=k\pi$
(k=0,1,2,...) and $\xi=1$: (a) $\Delta=0$ (dashed line); (b) $\Delta=1$ 
(dotted-dashed line);  (c) $\Delta=5$ (solid line).

\
 
{\bf FIG.7}: Time evolution of the function $F_{1}(P=6)$ for
$\eta=0.5$, $\theta=\pi/2$, $\phi=k\pi$
(k=0,1,2,...) and $\xi=3$: (a) $\Delta=0$ (dashed line); (b) $\Delta=1$
(dotted-dashed line);  (c) $\Delta=10$ (solid line).

\
 
{\bf FIG.8}:
The 3D contour plot of function $F_{2}(P=2)$,
$\phi$ and $\theta$ for $\Delta=0$,
$\sqrt{\xi!}\tau=k\pi$ and $\eta=\mid\sin\theta\cos\theta\mid$,
where the dotted line is for $F_{2}(P=2)=0$,
the dotted-dashed lines for $F_{2}(P=2) > 0$ in the range of
0.01 to 0.24 with an interval of 0.01, and the solid lines for
$F_{2}(P=2) < 0$ in the range of -0.06 to -0.005 with an interval
of 0.005.

\ 
 
{\bf FIG.9}:
 The 3D contour plot of function $F_{2}(P=2)$,
$\phi$ and $\theta$ for $\Delta=0$, $\sqrt{\xi!}\tau=k\pi$ and
$\eta=0.5$, where the dotted line is for $F_{2}(P=2)=0$,
the dotted-dashed lines for $F_{2}(P=2) > 0$ in the range of
0.01 to 0.22 with an interval of 0.03, and the solid lines for
$F_{2}(P=2) < 0$ in the range of -0.24 to -0.01 with an interval
of 0.01.

\newpage 

\begin{center}
\begin{center}{\Large\bf Xie \& Rao: FIG.1/PHYSICA A}\end{center}

\vspace{2cm}

\epsfig{file=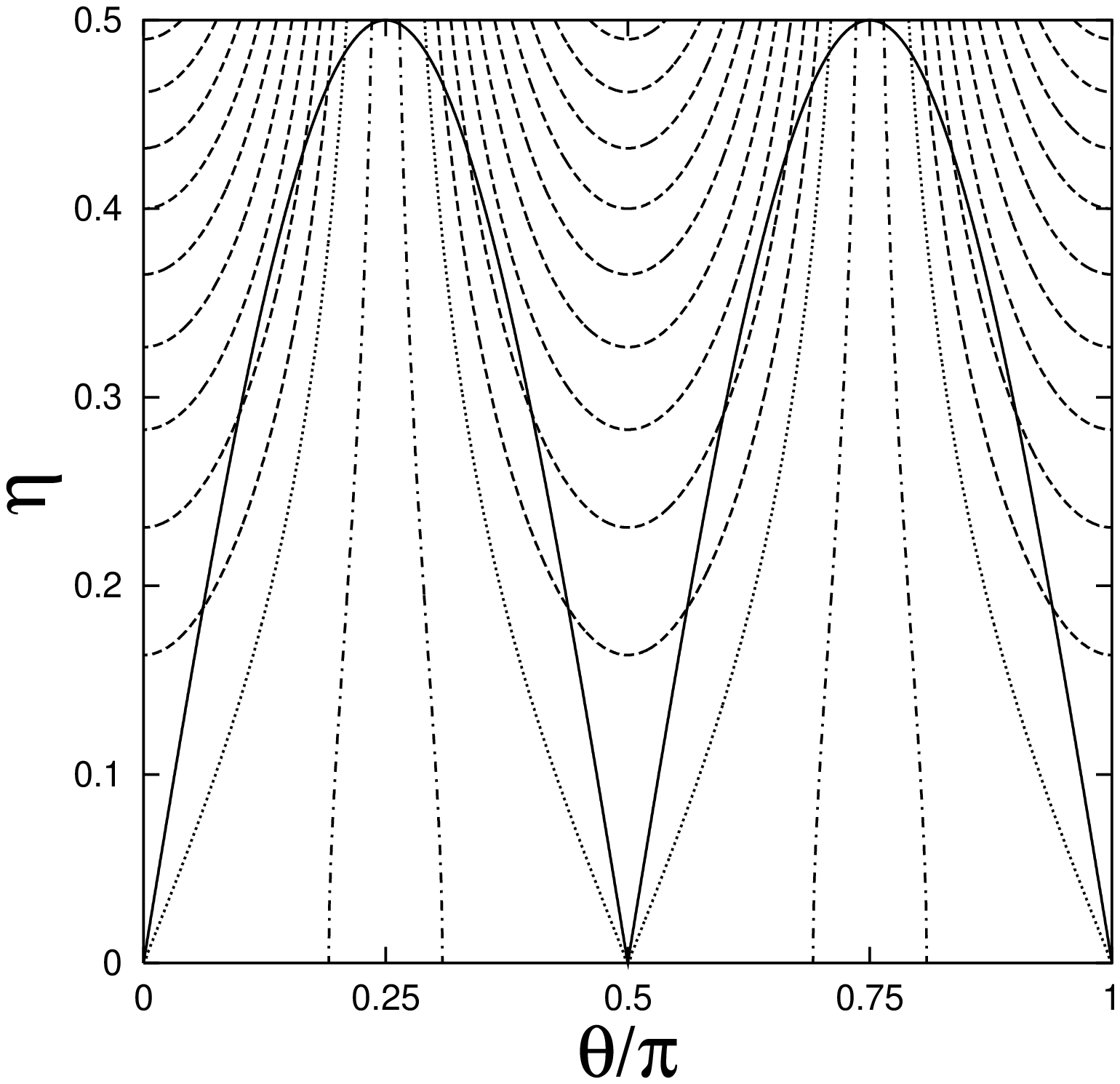}

\newpage

\begin{center}{\Large\bf Xie \& Rao: FIG.2/PHYSICA A}\end{center}
 
\vspace{2cm}
 
\epsfig{file=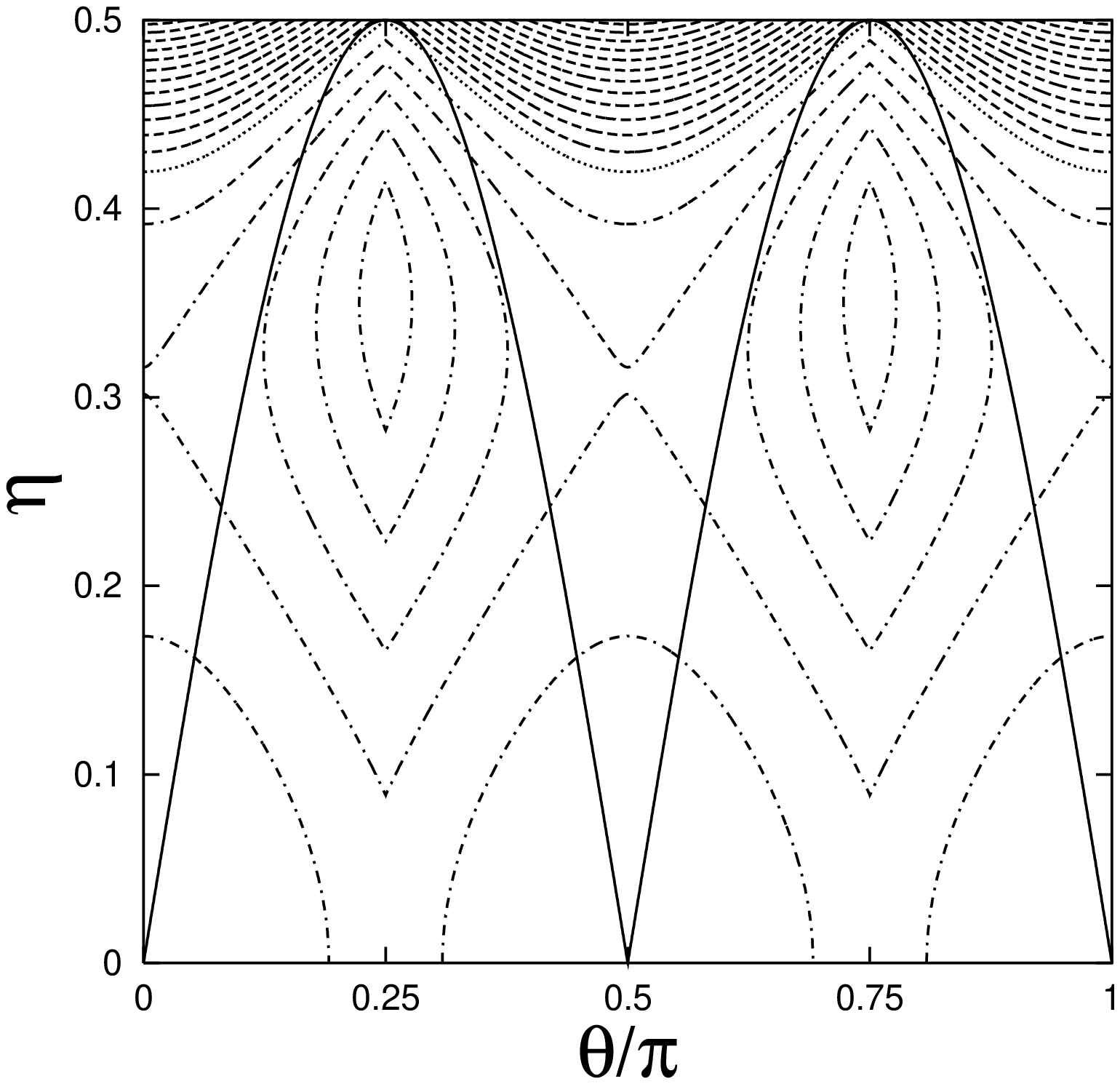}

\newpage

\begin{center}{\Large\bf Xie \& Rao: FIG.3/PHYSICA A}\end{center}
 
\vspace{2cm}
 
\epsfig{file=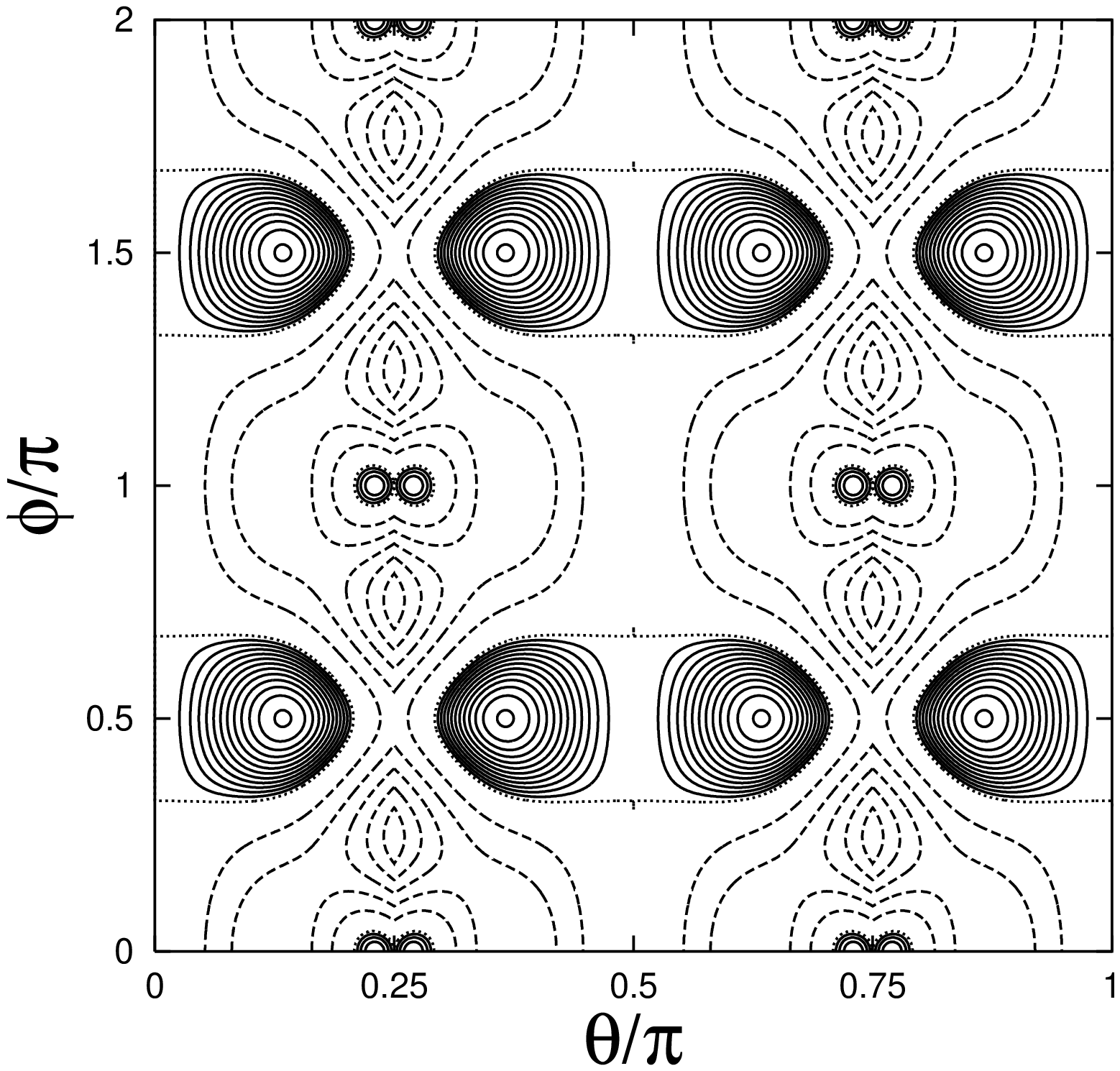}

\newpage 
 
\begin{center}{\Large\bf Xie \& Rao: FIG.4/PHYSICA A}\end{center}
 
\vspace{2cm}
 
\epsfig{file=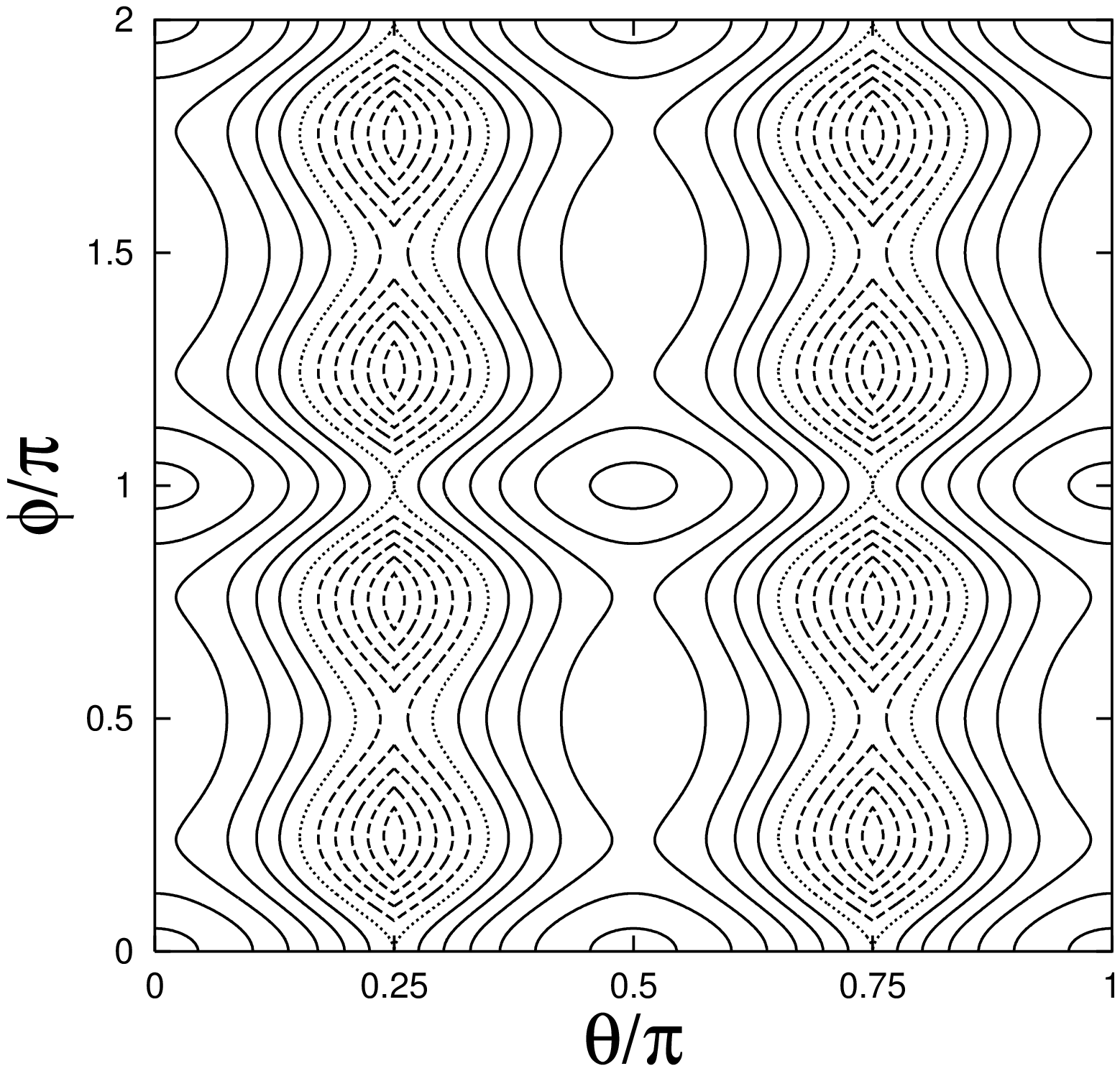}

\newpage
 
\begin{center}{\Large\bf Xie \& Rao: FIG.5/PHYSICA A}\end{center}
 
\vspace{2cm}
 
\epsfig{file=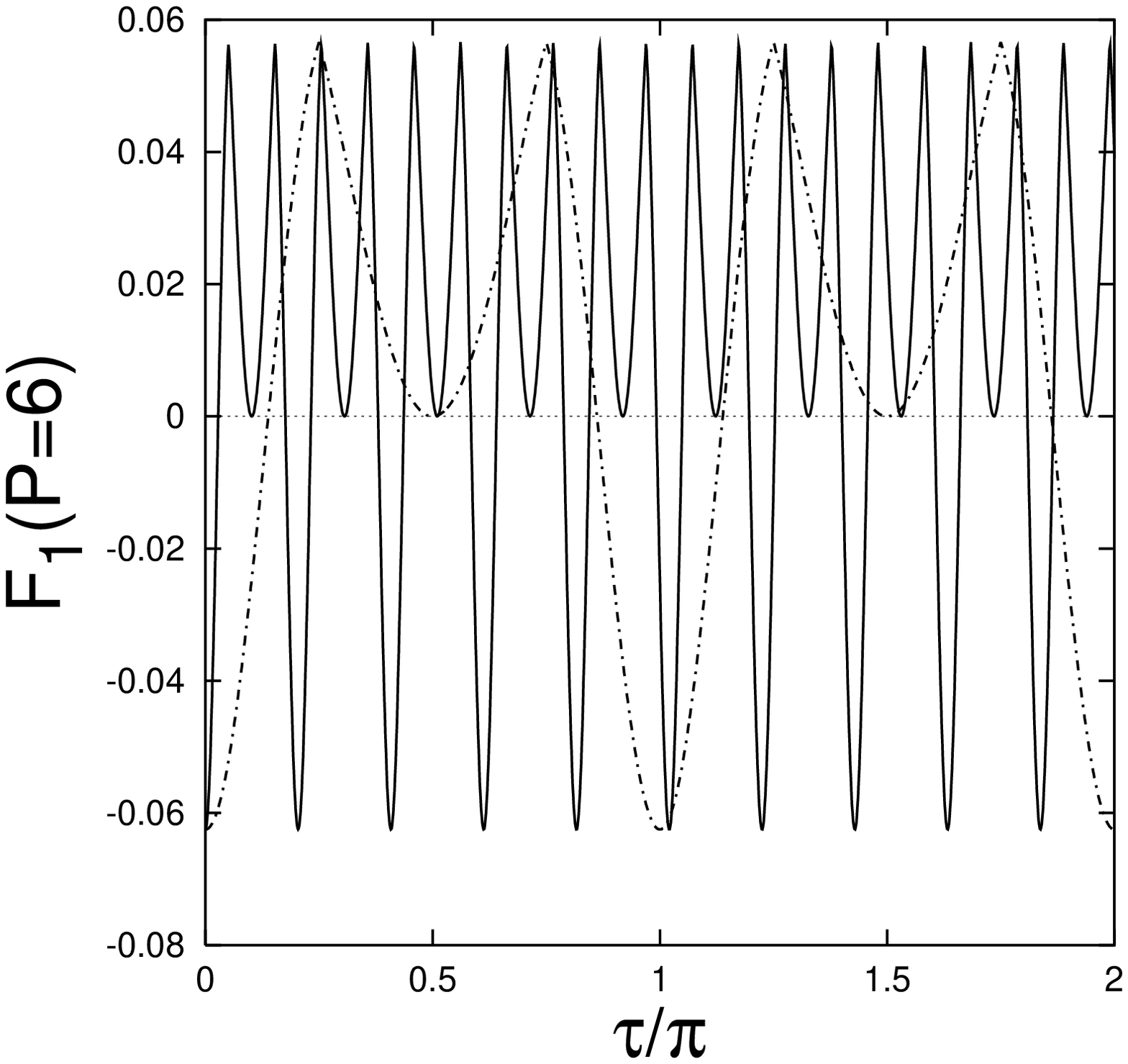}

\newpage
 
\begin{center}{\Large\bf Xie \& Rao: FIG.6/PHYSICA A}\end{center}
 
\vspace{2cm}
 
\epsfig{file=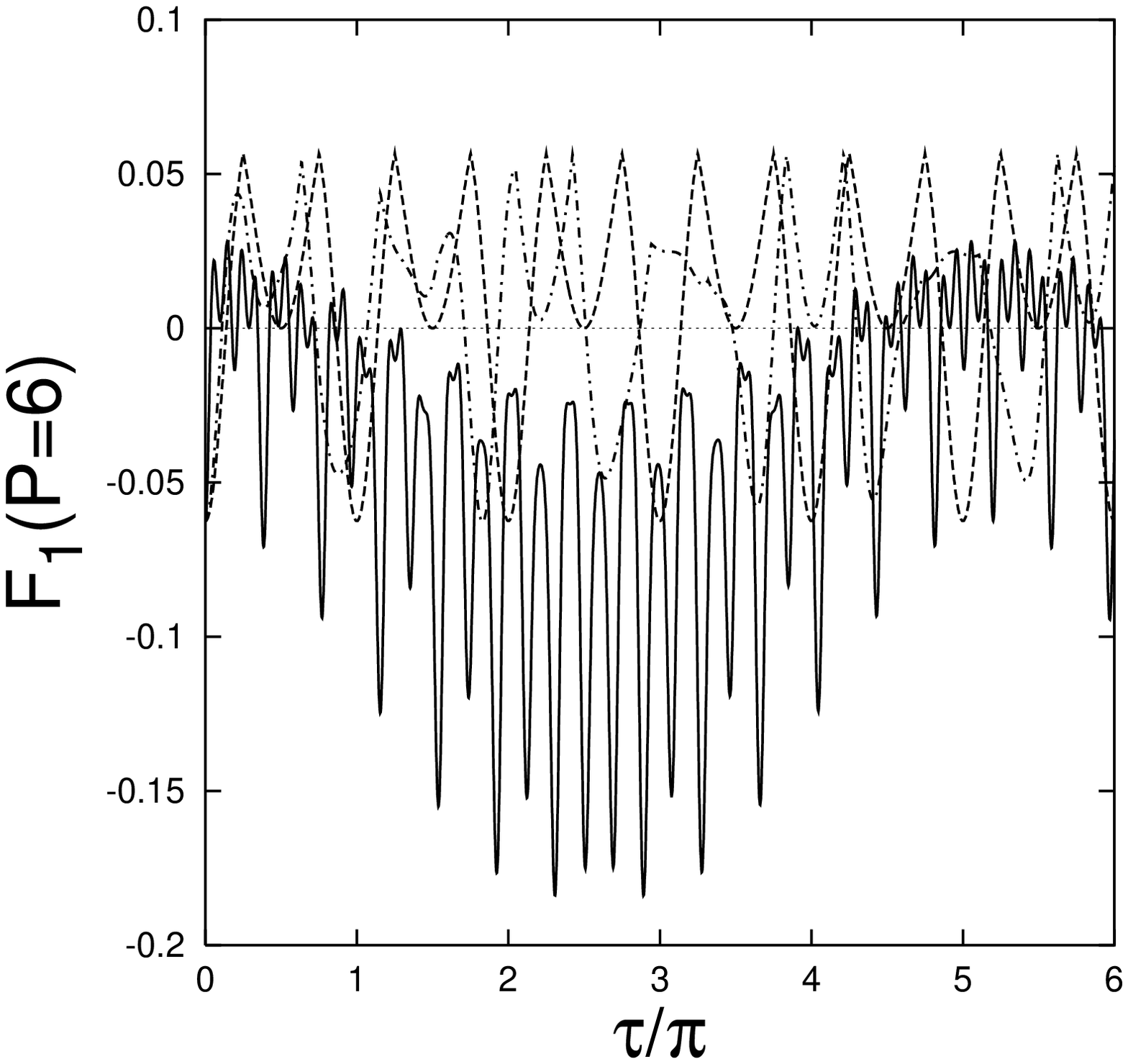}

\newpage
 
\begin{center}{\Large\bf Xie \& Rao: FIG.7/PHYSICA A}\end{center}
 
\vspace{2cm}
 
\epsfig{file=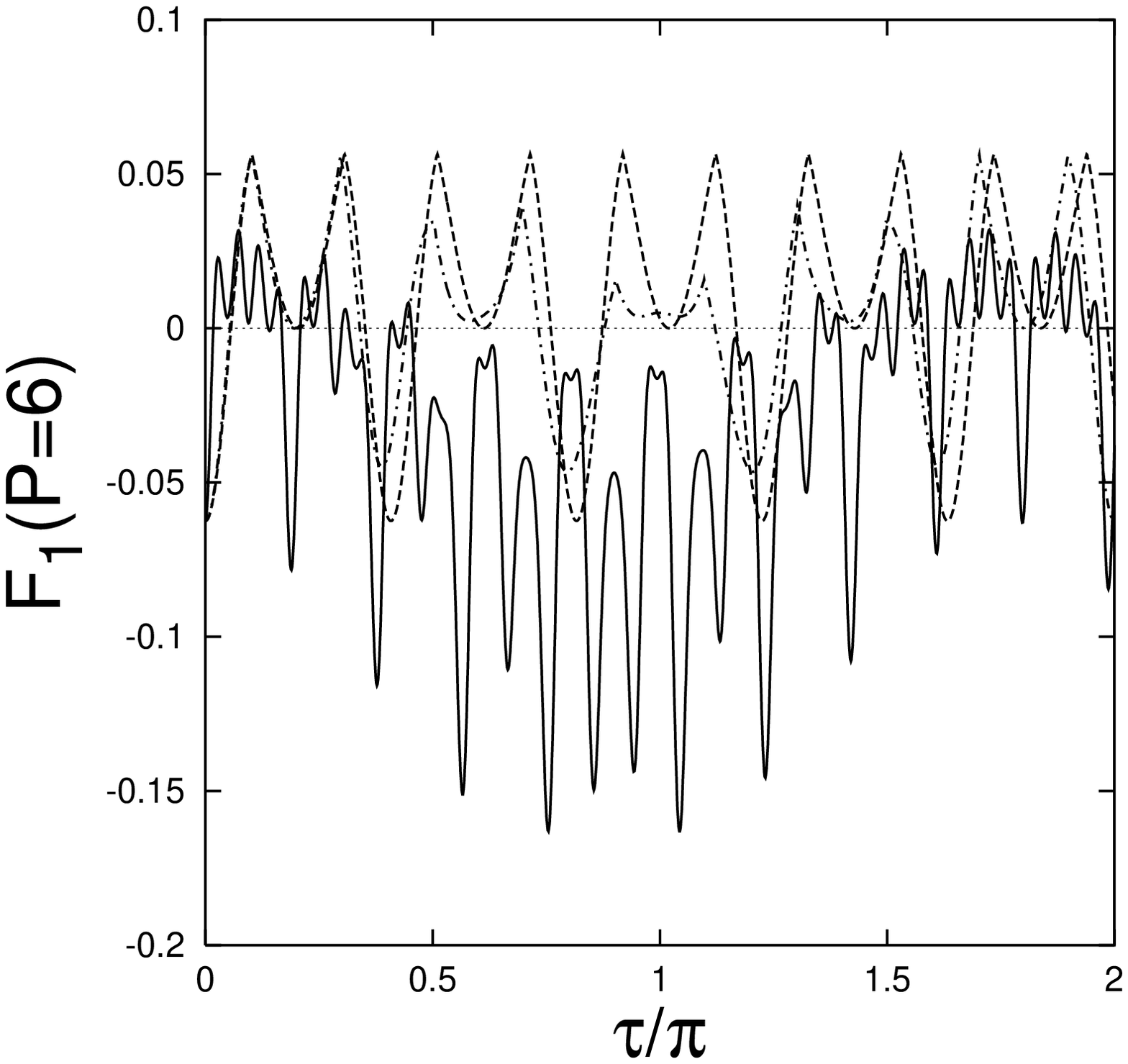}
\end{center}

\newpage
 
\begin{center}{\Large\bf Xie \& Rao: FIG.8/PHYSICA A}\end{center}
 
\vspace{2cm}
 
\epsfig{file=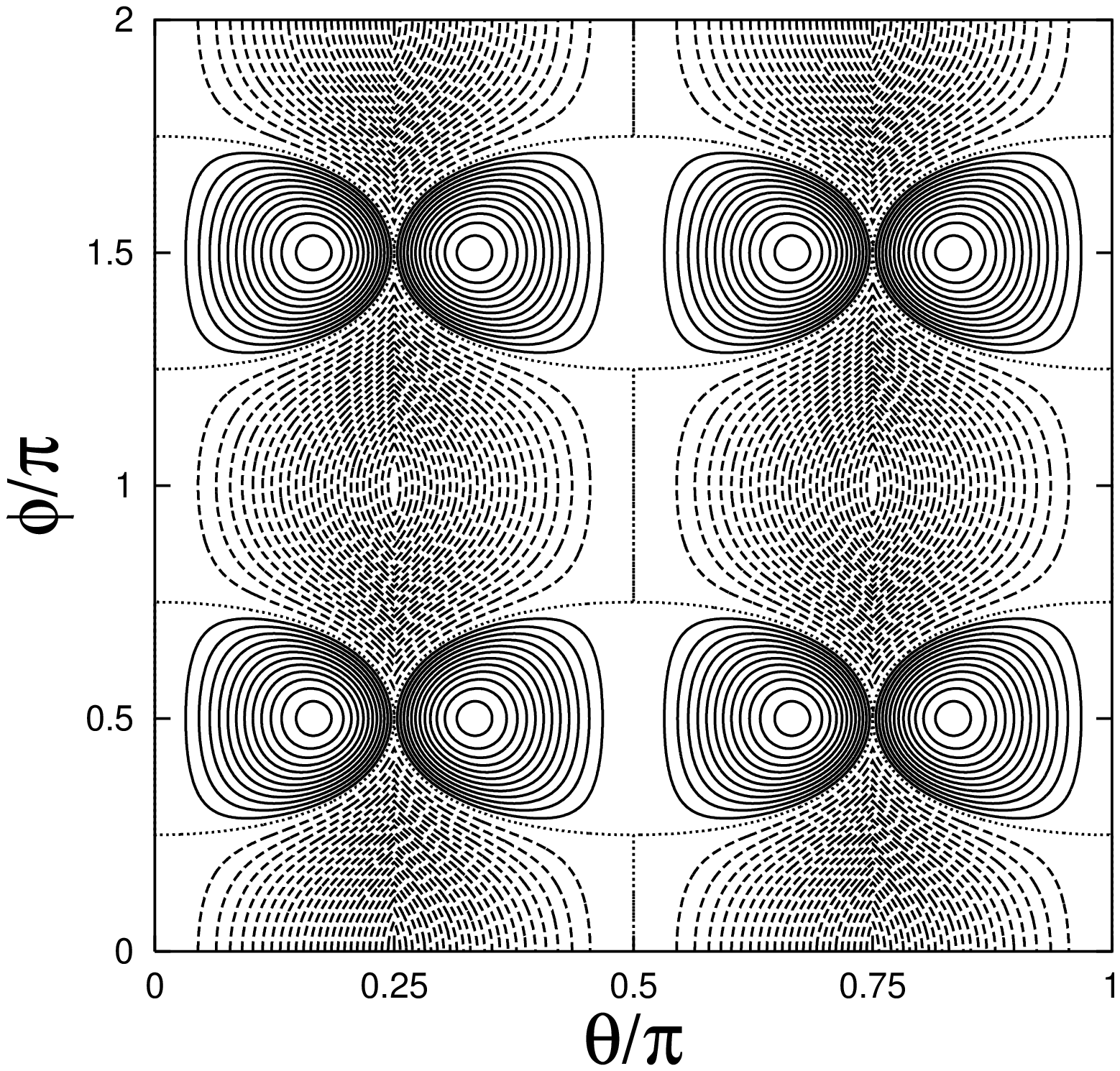}

\newpage
 
\begin{center}{\Large\bf Xie \& Rao: FIG.9/PHYSICA A}\end{center}
 
\vspace{2cm}
 
\epsfig{file=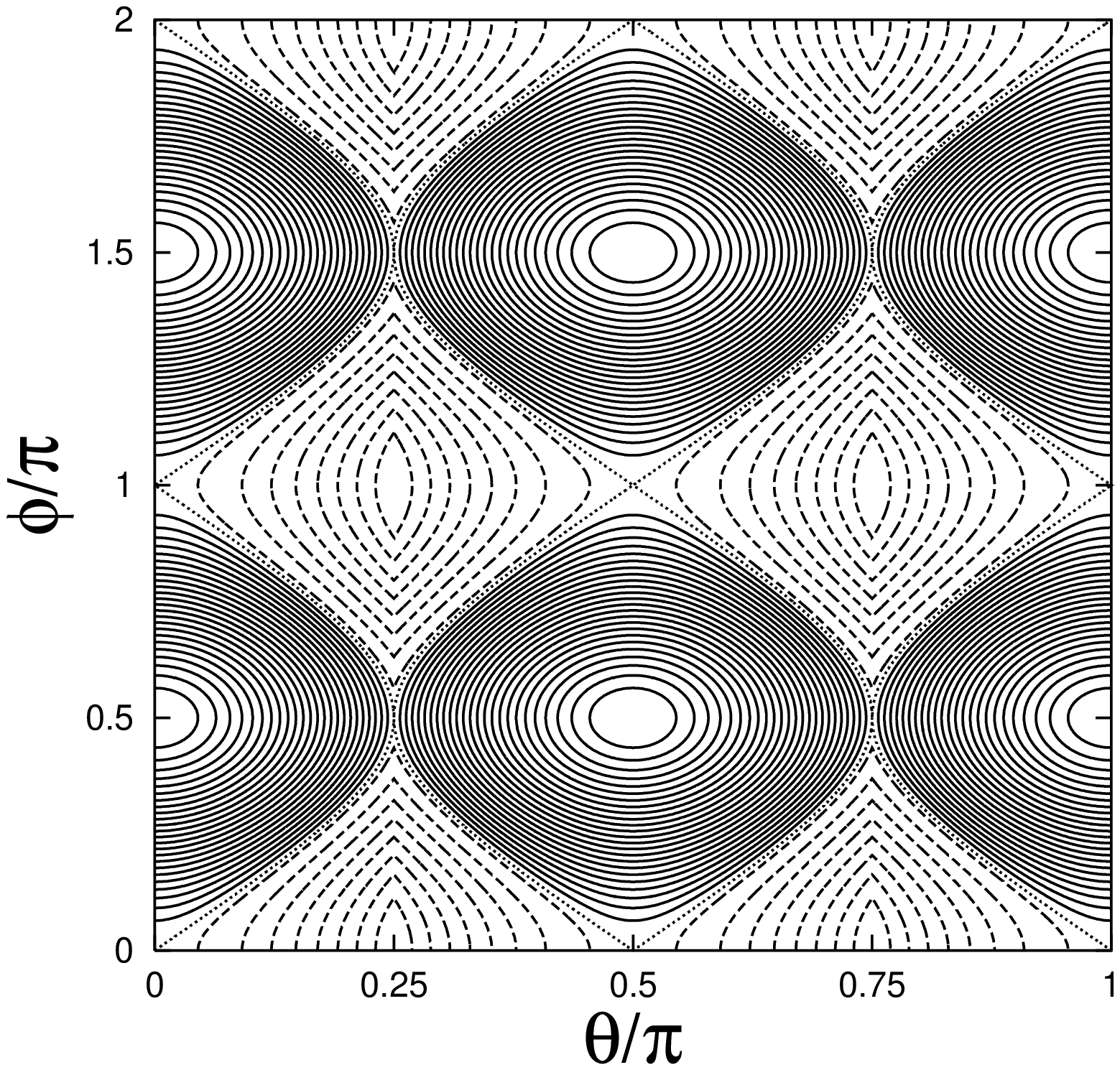}

\end{document}